\documentclass[pdflatex,sn-mathphys-num]{sn-jnl}% Math and Physical Sciences Numbered Reference Style
\usepackage{graphicx}%
\usepackage{multirow}%
\usepackage{amsmath,amssymb,amsfonts}%
\usepackage{amsthm}%
\usepackage{mathrsfs}%
\usepackage[title]{appendix}%
\usepackage{xcolor}%
\usepackage{textcomp}%
\usepackage{manyfoot}%
\usepackage{booktabs}%
\usepackage{algorithm}%
\usepackage{algorithmicx}%
\usepackage{algpseudocode}%
\usepackage{tabularx,booktabs,array}
\newcolumntype{L}{>{\raggedright\arraybackslash}X}
\usepackage{listings}%
\usepackage{multirow}
\usepackage{tabularray}
\UseTblrLibrary{siunitx}
\usepackage[listings,skins,breakable]{tcolorbox}
\usepackage{listings}
\lstdefinestyle{router}{
  basicstyle=\ttfamily\footnotesize,
  breaklines=true, breakatwhitespace=true,
  columns=fullflexible, keepspaces=true,
  showstringspaces=false
}
\theoremstyle{thmstyleone}%
\theoremstyle{thmstyletwo}%

\theoremstyle{thmstylethree}%

\begin{document}

\title[Article Title]{AT-CXR: Uncertainty-Aware Agentic Triage for Chest X-rays}

%%=============================================================%%
%% GivenName	-> \fnm{Joergen W.}
%% Particle	-> \spfx{van der} -> surname prefix
%% FamilyName	-> \sur{Ploeg}
%% Suffix	-> \sfx{IV}
%% \author*[1,2]{\fnm{Joergen W.} \spfx{van der} \sur{Ploeg} 
%%  \sfx{IV}}\email{iauthor@gmail.com}
%%=============================================================%%

% \author*[1,2]{\fnm{First} \sur{Author}}\email{iauthor@gmail.com}

% \author[2,3]{\fnm{Second} \sur{Author}}\email{iiauthor@gmail.com}
% \equalcont{These authors contributed equally to this work.}

% \author[1,2]{\fnm{Third} \sur{Author}}\email{iiiauthor@gmail.com}
% \equalcont{These authors contributed equally to this work.}

% \affil*[1]{\orgdiv{Department}, \orgname{Organization}, \orgaddress{\street{Street}, \city{City}, \postcode{100190}, \state{State}, \country{Country}}}

% \affil[2]{\orgdiv{Department}, \orgname{Organization}, \orgaddress{\street{Street}, \city{City}, \postcode{10587}, \state{State}, \country{Country}}}

% \affil[3]{\orgdiv{Department}, \orgname{Organization}, \orgaddress{\street{Street}, \city{City}, \postcode{610101}, \state{State}, \country{Country}}}
\author[1]{\fnm{Xueyang} \sur{Li}}\email{xli34@nd.edu}
\author[1]{\fnm{Mingze} \sur{Jiang}}\email{mjiang23@nd.edu}
\author[1]{\fnm{Gelei} \sur{Xu}}\email{gxu4@nd.edu}
\author[1]{\fnm{Jun} \sur{Xia}}\email{jxia4@nd.edu}
\author[1]{\fnm{Mengzhao} \sur{Jia}}\email{mjia2@nd.edu}
\author[1]{\fnm{Danny} \sur{Chen}}\email{dchen@nd.edu}
\author*[1]{\fnm{Yiyu} \sur{Shi}}\email{yshi4@nd.edu}

\affil*[1]{\orgname{University of Notre Dame}, 
\orgaddress{\city{Notre Dame}, \state{IN}, \country{USA}}}

\abstract{Agentic AI is advancing rapidly, yet truly autonomous medical-imaging triage, where a system decides when to stop, escalate, or defer under real constraints, remains relatively underexplored. To address this gap, we introduce \textbf{AT-CXR}, an uncertainty-aware agent for chest X-rays. The system estimates per-case confidence and distributional fit, then follows a stepwise policy to issue an automated decision or abstain with a suggested label for human intervention. We evaluate two router designs that share the same inputs and actions: a deterministic rule-based router and an LLM-decided router. Across five-fold evaluation on a balanced subset of NIH ChestX-ray14 dataset, both variants outperform strong zero-shot vision–language models and state-of-the-art supervised classifiers, achieving higher full-coverage accuracy and superior selective-prediction performance, evidenced by a lower area under the risk–coverage curve (AURC) and a lower error rate at high coverage, while operating with lower latency that meets practical clinical constraints. The two routers provide complementary operating points, enabling deployments to prioritize maximal throughput or maximal accuracy. Our code is available at \url{https://github.com/XLIAaron/uncertainty-aware-cxr-agent}.}


%%\pacs[JEL Classification]{D8, H51}

%%\pacs[MSC Classification]{35A01, 65L10, 65L12, 65L20, 65L70}

\maketitle

\section{Introduction}
Agentic AI systems, which select and sequence tools via a perceive-plan-act loop, have recently shown strong performance on reasoning and tool-use tasks~\cite{acharya2025agentic}. In the medical domain, agentic designs are increasingly explored for data curation, report drafting, and multimodal visual question answering (VQA)~\cite{zhang2025patho,shimgekar2025agentic,he2025medorch}. However, clinical deployment requires more than answer accuracy. In chest radiography, for example, a system must decide case-wise whether to accept and issue an automated label, escalate to additional and costlier analysis when uncertainty is high, or defer to a human reader under explicit cost and latency constraints. We refer to this decision problem as CXR triage, a sequential decision problem under uncertainty and resource constraints. Given workload and case complexity, clinicians increasingly seek AI assistance \cite{sridharan2024real}. However, traditional single-pass models act as black boxes, producing a score without governing analysis depth, deferral, or budget, and they provide little actionable uncertainty or shift awareness. By contrast, an agentic approach can organize triage as a policy over actions driven by uncertainty estimates and distributional fit, enabling selective automation with auditable operation.

To the best of our knowledge, the application of agentic AI to CXR triage has not been examined. Existing medical agentic frameworks emphasize diagnosis-oriented prediction, report drafting, or multimodal analysis, but these do not address the operational demands of triage. Direct application of existing medical agentic frameworks to CXR triage is problematic. Beyond differences in goals and overall structure, most prior designs neglect uncertainty estimation and distributional shift, and they do not incorporate these aspects into their decision-making policies, leaving systems vulnerable to being confidently wrong without mechanisms for self-assessment. Moreover, their computational complexity is substantial: many frameworks rely on large-scale models that require advanced GPUs such as the A6000~\cite{fallahpour2025medrax} or A100~\cite{li2024mmedagent} for deployment. Such requirements are incompatible with typical clinical environments, where deployment must be feasible on local workstations and operate with low latency. In addition, many existing agentic pipelines, such as VQA systems~\cite{he2025medorch}, still rely on physician prompts to initiate or guide the process. This reliance limits their degree of end-to-end automation and reduces suitability for real clinical workloads, where physicians cannot be expected to provide prompts for every case and efficiency is paramount.

To address these gaps, we propose \textbf{AT-CXR} (Agentic Triage for Chest X-ray), an \textbf{uncertainty-aware agentic framework} designed to make autonomous, safe triage decisions under clinical resource and reliability constraints. AT-CXR treats triage as a stepwise policy over a fixed action set: given an initial assessment score and a distribution-fit signal, the router may invoke test-time augmentation to probe stability, consult a committee of experts, or escalate to a vision–language model. The policy can iterate and then either issue an automated label or abstain with a suggested label for human review. A folder watcher and auto-sorting complete an end-to-end pipeline, reducing cognitive load and aligning with clinical throughput. For interpretability, AT-CXR generates Class Activation Maps~\cite{zhou2016learning} that highlight suspicious regions alongside the triage outcome. We instantiate two router designs over the same inputs and action set: a deterministic rule-based router and an LLM-decided router. Evaluated on NIH ChestX-ray14~\cite{wang2017chestx} for pulmonary-edema triage, the LLM and rule-based routers achieve overall accuracies of \textbf{95.3\%} and \textbf{93.8\%}, respectively, at 100\% coverage (including abstained-with-suggestion outputs). Both designs surpass strong zero-shot vision–language models and state-of-the-art supervised classifiers, achieving higher full-coverage accuracy and stronger selective-prediction performance, as reflected in lower AURC and reduced error at high coverage, while maintaining low latency and modest computational requirements that are compatible with clinical workflows. These results highlight the potential of uncertainty-aware agentic policies to enable safe and efficient CXR triage.

\section{Related Work}
In this section, we briefly review two areas most relevant to our study: (1) existing applications of agentic AI in the medical domain, which have primarily focused on diagnosis, report drafting, and multimodal question answering, and (2) vision–language models (VLMs), including both general-purpose and medical-specialized variants, which provide flexible multimodal reasoning but remain limited in calibration, efficiency, and end-to-end automation.
\subsection{Medical Agentic AI}
Agentic AI has gained traction in medical applications such as diagnostic decision-making, report generation, and visual question answering (VQA). A key direction involves multi-agent and tool-augmented diagnostic pipelines. MDAgents\cite{kim2024mdagents} configures LLM-based agents into teams based on task complexity. MedAgent-Pro\cite{wang2025medagent} combines guideline-driven planning with patient-level reasoning, integrating tools like imaging analyzers via RAG. MMedAgent-RL\cite{xia2025mmedagent} introduces reinforcement learning for dynamic role assignment and cooperative decision-making, improving performance in complex multimodal tasks.

Another direction explores tool-augmented single-agent systems. MedRAX\cite{fallahpour2025medrax} integrates chest X-ray tools with VLMs for structured reporting. MMedAgent\cite{li2024mmedagent} generalizes this approach across imaging modalities by learning to orchestrate tools like segmentation models and RAG-based modules. Additional systems incorporate knowledge retrieval and guideline-driven reasoning to enhance individual workflows.

Agentic Retrieval-Augmented Generation (Agentic RAG) has also emerged. Patho-AgenticRAG\cite{zhang2025patho} supports joint retrieval over text and image embeddings, using reinforcement learning for task decomposition and iterative search. Such frameworks are promising for domains requiring high-resolution and fine-grained reasoning.

Despite recent advances, current medical agentic frameworks remain limited for chest X-ray triage. Many VQA-style systems depend on physician-driven prompts, reducing automation potential. Most pipelines process cases independently, lacking cross-case workflow mechanisms such as diagnostic depth control, urgent-case prioritization, resource allocation, and latency optimization. Additionally, these frameworks focus on answer generation without incorporating uncertainty estimation or safe deferral strategies, which constrains reliability in high-risk settings. While uncertainty-aware agentic decision-making has been explored in non-medical domains\cite{zhi2025seeing}, its systematic application to clinical triage remains limited, hindering auditability, efficiency, and real-time performance.

\subsection{Vision Language Models}
General-purpose vision–language models (VLMs), such as GPT-4o\cite{openai_gpt4o_systemcard_2024}, GPT-4.1-mini\cite{openai_gpt41mini_docs_2025}, and GPT-5\cite{gpt5_introducing_2025}, demonstrate strong multimodal reasoning for tasks like diagnostic explanation and report interpretation. Their robust backend enables fast inference without requiring local computation. However, their proprietary and commercial nature introduces high costs and limits transparency, reproducibility, and on-premise deployment—key concerns in clinical environments.

Several VLMs have been fine-tuned for medical imaging. MedGemma-4b and MedGemma-27b\cite{sellergren2025medgemma} leverage large-scale radiology data and curated knowledge for improved domain-specific reasoning. LLaVA-Med\cite{li2023llava} enhances model transparency and deployability by incorporating expert-annotated medical images. Yet, these models remain designed for general medical imaging and are not specifically optimized for chest radiograph (CXR) interpretation.

CheXagent\cite{chen2024chexagent}, a CXR-focused VLM, integrates retrieval, reporting, and VQA components tailored for CXR interpretation. It enables localized image explanations and includes curated benchmarks. Nonetheless, it still relies on prompt-driven interactions and lacks support for fully automated triage workflows, limiting its effectiveness in high-throughput clinical environments.

In summary, existing systems emphasize diagnosis and reporting but fall short of supporting clinical triage. While VLMs provide flexible reasoning, they often lack uncertainty calibration, computational efficiency, and automation. Our proposed system, AT-CXR, directly addresses these gaps through uncertainty-aware decision-making, lightweight deployment, and end-to-end triage workflows.

\section{Methodology}
\begin{figure*}[t]
\centering
\includegraphics[width=\textwidth]{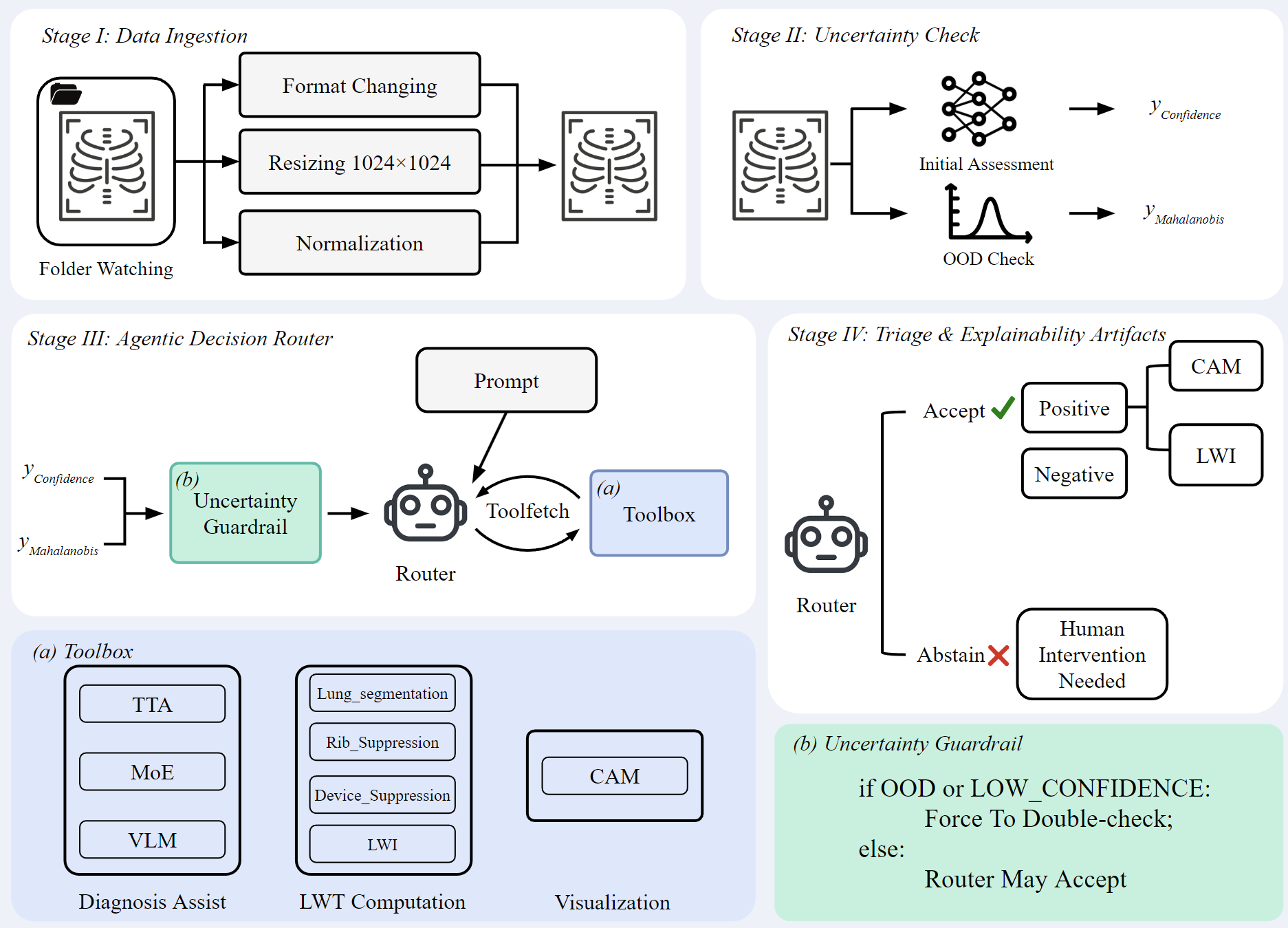}
\caption{Overview of \textbf{AT-CXR}. The framework comprises four stages: 
(I) \textit{Data Ingestion}, where a folder watcher detects new cases and triggers preprocessing; 
(II) \textit{Uncertainty Check}, which computes a confidence score \(y_{\text{confidence}}\) from a trained classifier and a distribution-fit signal \(y_{\text{Mahalanobis}}\) for out-of-distribution (OOD) assessment; 
(III) \textit{Agentic Decision Router}, which is guarded by an uncertainty guardrail, uses these signals and a task prompt to iteratively select tools from a fixed set until a stopping decision is reached;
(IV) \textit{Triage and Explainability Artifacts}, which auto-sorts cases (Positive/Negative) to corresponding folder or flags them for human intervention, and for accepted Positive cases generates Class Activation Maps~\cite{zhou2016learning} and a normalized lung whiteness index (LWI) computed within the lung fields after rib and medical-device suppression, as a proxy for edema extent. Insets: (a) \textit{Toolbox} grouped into diagnosis assist, LWI computation, and visualization; (b) \textit{Uncertainty Guardrail}, which forces a double-check when the case is OOD or low-confidence, and otherwise permits acceptance.}

\label{model1}
\end{figure*}

The general workflow of \textbf{AT-CXR} is depicted in Fig.~\ref{model1}. The system comprises four stages: (i) \textbf{Data Ingestion}, where a folder watcher detects new cases and automatically initiates processing; (ii) \textbf{Uncertainty Check}, which produces a baseline confidence estimate \(y_{\text{confidence}}\) from a trained classifier and a distribution-fit signal \(y_{\text{Mahalanobis}}\) for out-of-distribution assessment; (iii) \textbf{Agentic Decision Routing}, which executes a guardrailed, stepwise policy over a toolbox of actions and may iterate until a stopping criterion is satisfied; and (iv) \textbf{Triage and Explainability Artifacts}, which auto-sorts cases into outcome folders and, for accepted cases predicted as positive, generates Class Activation Maps~\cite{zhou2016learning} and computes a normalized lung whiteness index (LWI) within the lung fields after rib and medical-device suppression, as a proxy for edema extent. For the router design, we explore two types of policies that operate over the same inputs: a deterministic rule-based router and an LLM-decided router.

The following subsections provide an in-depth account of the \textbf{AT-CXR}, describing each stage and the underlying components.

\subsection{Stage I: Data Ingestion}\label{sec:ingestion}
An event-driven folder watcher monitors the input directory and automatically initiates processing when a new case appears. We accept DICOM as well as JPEG and PNG exports. DICOM cases are converted to PNG, and protected health information, where present, is removed from headers. Images are converted to 8-bit grayscale, resized to $1024\times1024$, and linearly scaled to $[0,1]$ by dividing pixel values by 255. Each case is assigned a unique identifier and enqueued with its metadata for the initial uncertainty check in Stage~II.

\subsection{Stage II: Uncertainty Check}\label{sec:stage2}
Stage~II computes two per-case signals on each preprocessed CXR from Stage~I: (a) a distribution–fit score that quantifies how well the case conforms to a reference feature distribution, and (b) a baseline classifier confidence for the target label.

\subsubsection{Out-of-distribution (OOD) Check via Mahalanobis}
We use the Mahalanobis distance~\cite{mahalanobis2018generalized} in a standardized radiomics feature space as the OOD score; this serves as the distributional-fit signal consumed by the router. The method is principled under a Gaussian reference model, relies only on reference-set summary statistics (mean and covariance), and is computationally lightweight, requiring no additional training or prompting. These properties align with clinical latency and deployability constraints.

% For cohort-level monitoring (not used by the router), we additionally compute a Fréchet distance (FRD) between Gaussian fits of feature sets.

Let \(\mathcal{D}_{\text{ref}}=\{x_i\}_{i=1}^{n}\) denote a fixed reference dataset (e.g., the training split).
For each preprocessed CXR \(x_i\), we extract a radiomics feature vector using PyRadiomics~\cite{vanGriethuysen2017Computational} (2D, Original image type, binWidth=10) on a whole-image mask with a one-pixel zero border to define background. Let \(\psi(x_i)\in\mathbb{R}^{D_{\text{raw}}}\) denote the vector of all radiomics features returned by PyRadiomics for image \(x_i\) when using the \texttt{Original} image type (i.e., features computed on the unfiltered image; examples include \texttt{original\_firstorder\_*}, \texttt{original\_glcm\_*}, \texttt{original\_glrlm\_*}, \texttt{original\_glszm\_*}, \texttt{original\_gldm\_*}, \texttt{original\_ngtdm\_*}). After stacking \(\{\psi(x_i)\}_{i=1}^n\), we (i) impute occasional missing values, (ii) remove columns that are identically zero across the stack, and (iii) align the remaining features to a fixed name/column order. Denote by \(\mathcal{K}\subseteq\{1,\dots,D_{\text{raw}}\}\) the index set of retained features and let \(d=|\mathcal{K}|\). We then define the aligned feature vector
\[
\phi(x_i)\;=\;\psi(x_i)_{\mathcal{K}}\in\mathbb{R}^{d},
\]
which ensures a consistent feature space across all cases for subsequent standardization and covariance modeling. On this aligned feature space, we compute the standardized feature for each reference image, $z_i$, as
\[
z_i \;=\; \mathrm{Std}\!\big(\phi(x_i)\big)
\;=\; \frac{\phi(x_i)-\mu^{\text{std}}}{\sigma^{\text{std}}},
\qquad x_i\in\mathcal{D}_{\text{ref}},
\]
in which \(\mu^{\text{std}}\) and \(\sigma^{\text{std}}\) are the per-feature mean and standard deviation fitted on \(\{\phi(x_i)\}_{i=1}^{n}\).
We then estimate the reference mean and covariance in the standardized feature space:
\[
\mu_{\text{ref}} \;=\; \frac{1}{n}\sum_{i=1}^{n} z_i,
\qquad
\Sigma_{\text{ref}} \;=\; \frac{1}{n-1}\sum_{i=1}^{n} \big(z_i-\mu_{\text{ref}}\big)\big(z_i-\mu_{\text{ref}}\big)^{\!\top}.
\]
Here, \(\mu_{\text{ref}}\) is the empirical mean of the standardized radiomics features over \(\mathcal{D}_{\text{ref}}\), and \(\Sigma_{\text{ref}}\) is the corresponding empirical covariance matrix.

% \paragraph{Single-image distribution fit (Mahalanobis).}
Given a new input CXR image \(x\), we form its aligned, standardized feature
\(z=\mathrm{Std}\!\big(\phi(x)\big)\) and measure its fit to the reference distribution \(\mathcal{N}(\mu_{\text{ref}},\Sigma_{\text{ref}})\) via the Mahalanobis distance:
\[
y_{\text{Mahalanobis}}
\;=\;
\sqrt{\,\big(z-\mu_{\text{ref}}\big)^{\!\top}\,\Sigma_{\text{ref}}^{-1}\,\big(z-\mu_{\text{ref}}\big)\,}.
\]
% For numerical stability prior to inversion, we symmetrize and lightly regularize the covariance,
% \(\Sigma_{\text{ref}}\leftarrow \tfrac{1}{2}\big(\Sigma_{\text{ref}}+\Sigma_{\text{ref}}^{\top}\big)+\varepsilon I\) with a small \(\varepsilon>0\). We also record the Euclidean deviation \(\lVert z-\mu_{\text{ref}}\rVert_2\) as an auxiliary indicator. (For cohort-level monitoring between datasets, we additionally compute a Fréchet distance between Gaussian fits; see Appendix~\S\ref{app:frd} for details.)

\subsubsection{Initial Assessment}
In parallel with the OOD check, a supervised classifier provides a baseline confidence for each case. Given its favorable speed–accuracy trade-off on pulmonary edema, we use RexNet-150~\cite{han2021rethinking} in our design. The network outputs the estimated posterior \(P(y=\mathrm{edema}\mid x)\), implemented as a two-logit softmax; we denote this by \(y_{\text{confidence}}\in[0,1]\). Together with the Mahalanobis OOD score \(y_{\text{Mahalanobis}}\), this yields the uncertainty input \((y_{\text{confidence}},\,y_{\text{Mahalanobis}})\) to the Stage~III router.

\subsection{Stage III: Agentic Decision Router}
\subsubsection{Uncertainty Guardrail}
As presented in Fig.~\ref{model1}(b), we gate the router’s action space using the pair \((y_{\text{confidence}},\,y_{\text{Mahalanobis}})\).
Let \(\tau_{\text{OOD}}\) be the \(95^{\text{th}}\) percentile of the Mahalanobis scores computed on the reference set; a case is flagged out-of-distribution (OOD) if \(y_{\text{Mahalanobis}}>\tau_{\text{OOD}}\).
Let \(\tau_{\text{confidence}}\in(0,1)\) denote a high-confidence threshold selected on a held-out set to meet a desired operating point.

The guardrail enforces:
(i) if OOD (\(y_{\text{Mahalanobis}}>\tau_{\text{OOD}}\)) or not high-confidence (\(y_{\text{confidence}}<\tau_{\text{Confidence}}\)), the router is not permitted to issue a direct \textit{accept}; it must first invoke a verification/consultation tool (e.g., \texttt{TTA}, \texttt{MoE}, or \texttt{VLM}) or \textit{abstain};
(ii) a direct \textit{accept} (without \texttt{TTA}/\texttt{MoE} double-checks) is permitted only when the case is in-distribution and high-confidence, i.e., \(y_{\text{Mahalanobis}}\le \tau_{\text{OOD}}\) and \(y_{\text{confidence}}\ge \tau_{\text{Confidence}}\).
This guardrail is deterministic and model-agnostic, and it constrains the router’s policy rather than replacing it.

\begin{algorithm}[H]
% \caption{\textsc{UncertaintyGuardrail}$(y_{\text{confidence}}, y_{\text{Mahalanobis}}, \tau_{\text{confidence}}, \tau_{\text{OOD}})$}
\caption{\textsc{UncertaintyGuardrail}}
\label{alg:guardrail}
\begin{algorithmic}[1]
\Require $y_{\text{confidence}}\in[0,1]$, $y_{\text{Mahalanobis}}\ge 0$, thresholds $\tau_{\text{confidence}}$, $\tau_{\text{OOD}}$
\State $\text{OOD} \gets (y_{\text{Mahalanobis}} > \tau_{\text{OOD}})$
\If{$\text{OOD}$ \textbf{or} $y_{\text{confidence}} < \tau_{\text{confidence}}$}
    \State $\text{allowed\_actions.allow\_accept} \gets \textbf{False}$ \Comment{no direct \texttt{accept}}
\Else
    \State $\text{allowed\_actions.allow\_accept} \gets \textbf{True}$ \Comment{direct \texttt{accept} permitted}
\EndIf
\State \Return \text{allowed\_actions}, \text{OOD}
\end{algorithmic}
\end{algorithm}

\begin{table*}[ht]
\caption{AT-CXR toolbox: functions and policy effects. Lower-cost tools are preferred first (TTA $<$ MoE $<$ VLM). Thresholds ($\tau_{\mathrm{TTA}}$, $\tau_{\mathrm{MoE}}$, $\tau_{\mathrm{VLM}}$, $\tau_{\mathrm{Confidence}}$) are selected on a held-out set.}
\label{tab:toolbox}
\small
\begin{tblr}{
  hlines, vlines,
  colspec = {Q[c,m,9em] Q[l,m] X[l,m] X[l,m]},
  colsep  = 4pt,
  row{1}  = {guard, font=\bfseries, bg=gray!30}
}
Category & Tool name & Function & Policy effect \\

\SetCell[r=3]{c} Diagnosis assist
  & TTA &
    $K$ benign augmentations; compute $\sigma_{\mathrm{TTA}}$ of $P(y{=}\mathrm{edema}\mid x)$. &
    If $\sigma_{\mathrm{TTA}}\le \tau_{\mathrm{TTA}}$, set \texttt{allow\_accept} $\leftarrow$ \textbf{True}; else unchanged. \\
  & MoE &
    $E$ experts; majority vote and agreement rate $a$. &
    If $a\ge \tau_{\mathrm{MoE}}$, set \texttt{allow\_accept} $\leftarrow$ \textbf{True}; else unchanged. \\
  & VLM &
    Escalation under a structured prompt; returns label and rationale. &
    \textbf{Terminal:} issue \texttt{abstain} with suggested label and rationale; no automated \texttt{accept} after a VLM call. \\

\SetCell[r=4]{c} LWI computation
  & Lung segmentation &
    Segment lungs. &
    Post-accept only. \\
  & Rib suppression &
    Segment ribs and mask them. &
    Post-accept only. \\
  & Device suppression &
    Segment devices and mask them. &
    Post-accept only. \\
  & Lung whiteness index (LWI) &
    Mean intensity in lung area, with rib and medical devices suppressed. &
    Post-accept only. \\

Visualization
  & CAM &
    Class activation map from the base classifier. &
    Post-accept only. \\
\end{tblr}
\end{table*}

\subsubsection{Router and Toolbox}
The router, an LLM-decided policy using GPT\mbox{-}4.1\mbox{-}mini, receives the allowed-action set from the uncertainty guardrail and selects tools iteratively until a stopping decision (\textit{accept} or \textit{abstain}). Prompts for the LLM router is provided in the Appendix~\ref{AppA}. As illustrated in Fig.~\ref{model1}(a) and summarized in Table~\ref{tab:toolbox}, the toolbox comprises three categories, detailed below.

\textit{Diagnosis assistance.}
\texttt{TTA} applies the base RexNet\mbox{-}150 to \(K\) benign augmentations of the same image (e.g., small flips, rotations, contrast jitter), producing posterior samples whose mean \(\mu_{\mathrm{TTA}}\) and standard deviation \(\sigma_{\mathrm{TTA}}\) quantify stability; if \(\sigma_{\mathrm{TTA}}\le \tau_{\mathrm{TTA}}\) and \(\mu_{\mathrm{TTA}}\ge \tau_{\mathrm{Confidence}}\), the router may proceed to \textit{accept}. 
\texttt{MoE} queries a four-model committee (RexNet\mbox{-}150, RegNetY\mbox{-}1.6~\cite{radosavovic2020designing}, ResNet\mbox{-}50~\cite{he2016deep}, EfficientNet\mbox{-}V2~\cite{tan2021efficientnetv2}) and aggregates by majority vote; if the agreement rate \(a\ge \tau_{\mathrm{MoE}}\), \textit{accept} becomes permissible. 
\texttt{VLM} is a high-cost escalation that uses a structured prompt (GPT\mbox{-}4.1\mbox{-}mini with vision) to return a label and rationale; given its comparatively lower zero-shot accuracy on this task relative to trained classifiers, we treat the VLM as a terminal action that issues \textit{abstain} with a suggested label, providing a safety-preserving endpoint when verification does not yield sufficient consensus. The VLM diagnostic prompt is provided in the Appendix~\ref{AppB}.

\textit{LWI computation.}
For cases accepted as positive, the router triggers a post-accept pipeline that estimates parenchymal opacity while minimizing confounding from osseous structures and medical devices; to this end, we first segment these structures and suppress their signal before quantification.
First, a \texttt{lung\_segmentation} model produces a binary mask \(M_{\mathrm{lung}}\).
Next, \texttt{rib\_segmentation\_and\_suppression} and \texttt{device\_segmentation\_and\_suppression} both generate masks and actively remove their contributions from the image used for measurement.
Let \(M_{\mathrm{rib}}\) and \(M_{\mathrm{dev}}\) denote the rib and device masks, respectively, and let \(I:\Omega\!\to\![0,1]\) be the normalized grayscale image.
Rib suppression replaces intensities within \(M_{\mathrm{rib}}\) via a border-adaptive fill (each masked pixel is assigned the median intensity of a thin surrounding annulus), preserving local statistics without sharp seams.
Device suppression is performed by learned inpainting with LaMa~\cite{suvorov2022resolution} on \(M_{\mathrm{dev}}\), yielding a structurally plausible completion.
Denote the composed suppression by
\[
\tilde{I} \;=\; \mathcal{S}_{\mathrm{dev}}\!\big(\,\mathcal{S}_{\mathrm{rib}}(I,\,M_{\mathrm{rib}}),\,M_{\mathrm{dev}}\,\big),
\]
where \(\mathcal{S}_{\mathrm{rib}}\) is the border-adaptive rib fill and \(\mathcal{S}_{\mathrm{dev}}\) is LaMa inpainting under the device mask.
We then compute the \textbf{lung whiteness index (LWI)} on the suppressed image within the lung fields:
\[
\mathrm{LWI} \;=\; \frac{1}{\lvert M_{\mathrm{lung}}\rvert}\sum_{p\in M_{\mathrm{lung}}} \tilde{I}(p) \;\in\; [0,1].
\]
LWI is the mean post-suppression intensity over the lungs; higher values indicate greater parenchymal opacity and provide a simple, auditable scalar summary.
Although LWI targets opacity quantification, the segmentation and suppression operators \(\{M_{\mathrm{lung}}, M_{\mathrm{rib}}, M_{\mathrm{dev}}\}\) are pathology-agnostic and can be reused for other chest-radiograph tasks or coupled with alternative quantification modules.

\textit{Class activation map.}
To complement the opacity summary and support interpretability, we generate a Class Activation Map (CAM) from the base classifier, providing a qualitative localization cue for reviewer assessment.
For cases accepted as positive, we compute a gradient-based CAM at the final convolutional block for the predicted class, upsample it to the input resolution, normalize to \([0,1]\), and overlay it as a heatmap on the radiograph; both the raw CAM and the overlaid image are saved alongside the decision.

\subsubsection{Deterministic Rule-Based Router}
To assess the added value of the LLM-decided policy, we also implement a deterministic router that operates over the same inputs, thresholds, and action costs. 
Subject to the guardrail, the policy proceeds in a fixed order: \texttt{TTA} \(\rightarrow\) \texttt{MoE} \(\rightarrow\) \texttt{VLM}. 
If TTA satisfies \(\sigma_{\mathrm{TTA}}\le \tau_{\mathrm{TTA}}\) and \(\mu_{\mathrm{TTA}}\ge \tau_{\mathrm{Confidence}}\), the case is \textit{accepted}; otherwise the router queries the MoE and \textit{accepts} only if the agreement rate \(a\ge \tau_{\mathrm{MoE}}\). 
If neither criterion is met, the router escalates to the VLM, which is treated as a terminal \textit{abstain} with a suggested label. When a case is accepted as positive, the subsequent opacity quantification and visualization modules are also invoked.
Thus, both variants share inputs and thresholds; the difference lies solely in action selection -- adaptive (LLM-decided) versus fixed-order (rule-based).

\subsection{Stage IV: Triage and Explainability Artifacts}
Given the router’s stopping decision, the system performs result triage and persists artifacts for audit and clinical review. If the case is \textit{accepted}, it is automatically routed to a destination folder named by the final label (e.g., \texttt{positive} or \texttt{negative}). If the case is \textit{abstained}, it is placed in a dedicated \texttt{Human\_Intervention\_Needed} queue together with a suggested label and rationale.

For accepted positive cases, the pipeline saves (i) a class activation map (CAM) and its overlay on the radiograph, and (ii) the scalar lung–whiteness index (LWI). For all outcomes -- accepted or abstained -- the system writes a machine-readable trace (JSON) containing the case identifier, timestamps, model and tool versions, thresholds, uncertainty signals \((y_{\text{confidence}}, y_{\text{Mahalanobis}})\), the sequence of tools invoked with their key outputs, and the final decision with the deciding agent. This artifact bundle enables reproduction, auditing, and efficient human review.

\section{Experiments}

We evaluate \textbf{AT-CXR} on pulmonary-edema (PE) triage as a case study. Experiments use a class-balanced subset of NIH ChestX-ray14~\cite{wang2017chestx} containing 1,000 frontal radiographs (500 PE-positive, 500 PE-negative). The processed subset is publicly available on Kaggle.\footnote{\url{https://www.kaggle.com/datasets/samiulbari/pulmonary-edema-classified-by-nih/data}} Operating thresholds are as follows: a high-confidence threshold $\tau_{\mathrm{conf}}=0.60$; an OOD threshold $\tau_{\mathrm{OOD}}$ defined as the 95th percentile of Mahalanobis scores on the reference set; a TTA stability criterion $\tau_{\mathrm{TTA}}=0.05$ on the posterior standard deviation; and a mixture-of-experts (MoE) agreement threshold $\tau_{\mathrm{MoE}}=0.75$. Both router variants (LLM-decided and rule-based) use the same thresholds and receive identical inputs.

For baseline selection, we evaluate two groups: (i) zero-shot VLMs, comprising general-purpose models (GPT-4o~\cite{openai_gpt4o_systemcard_2024}, GPT-4.1~\cite{openai_gpt41_2025}, GPT-4.1-mini~\cite{openai_gpt41mini_docs_2025}, GPT-5~\cite{gpt5_introducing_2025}, GPT-5-mini~\cite{gpt5_introducing_2025}) and medical-specialized models (MedGemma-4B/27B~\cite{sellergren2025medgemma}, CheXagent~\cite{chen2024chexagent}, MedRAX~\cite{fallahpour2025medrax}, LLaVA-Med~\cite{li2023llava}); and (ii) supervised classifiers trained on the same data: RegNetY-1.6~\cite{radosavovic2020designing}, ResNet-50~\cite{he2016deep}, EfficientNet-V2~\cite{tan2021efficientnetv2}, and RexNet-150~\cite{han2021rethinking} (also used for the Stage~II initial assessment). For VLMs we use a standardized radiology prompt consistent with our router prompt but with orchestration instructions removed; full prompts are provided in the Appendix~\ref{AppB}. Supervised models are trained with 5-fold stratified cross-validation using identical splits; we report fold-averaged metrics. To reflect typical clinical workstations, all runs use a single NVIDIA A10 GPU without model parallelism.

Our primary endpoint is overall accuracy at 100\% coverage. For AT-CXR, every case receives a final output: automated \textit{accept} or \textit{abstain-with-suggested-label}. We also report precision and recall at 100\% coverage. For methods that expose a confidence score (AT-CXR and supervised baselines), we quantify selective automation with the Area Under the Risk–Coverage curve (AURC), risk@80\% coverage, and coverage@5\% risk, where coverage is the fraction of cases auto-resolved and risk is the error rate on that auto-resolved subset. Finally, we measure end-to-end per-case latency from case detection to final output, including any tool invocations. 

\begin{table*}[t]
\centering
\caption{Full-coverage (\(100\%\)) classification results for pulmonary-edema triage. Metrics are reported as mean~\(\pm\)~standard deviation over five cross-validation folds for Accuracy, Precision, and Recall at a fixed decision threshold of \(0.5\). All methods are evaluated on the same splits; bold indicates the best value in each column. Under the full-coverage setting, AT-CXR produces a final label for every case (accept or abstain-with-suggested-label).}
\label{table:result}
\setlength{\tabcolsep}{3.5pt}
\footnotesize
\resizebox{\textwidth}{!}{%
\begin{tabular}{llccc}
\toprule
\textbf{Category} & \textbf{Model Name} & \textbf{Accuracy (\%)} & \textbf{Precision (\%)} & \textbf{Recall (\%)}\\
\midrule
\multirow{9}{*}{Zero-Shot VLMs}
& GPT-4o            & 73.4 $\pm$ 2.3 & 80.0 $\pm$ 3.5 & 70.7 $\pm$ 3.5   \\
& GPT-4.1           & 78.2 $\pm$ 2.3 & 84.7 $\pm$ 3.9 & 75.0 $\pm$ 3.1   \\
& GPT-4.1-mini      & 79.8 $\pm$ 2.8 & 84.6 $\pm$ 4.7 & 77.1 $\pm$  3.0  \\
& GPT-5             & 80.3 $\pm$ 1.6 & 78.4 $\pm$ 3.1 & 83.8 $\pm$ 3.1   \\
& GPT-5-mini        & 80.3 $\pm$ 3.6 & 79.3 $\pm$ 4.7 & 81.0 $\pm$ 3.9   \\
& MedGemma-4b       & 75.3 $\pm$ 5.0 & 62.5 $\pm$ 8.3 & 83.8 $\pm$ 4.2   \\
& MedGemma-27b      & 81.2 $\pm$ 3.9  & 77.9 $\pm$ 4.2 & 83.7 $\pm$ 4.6   \\
& CheXagent       & 77.6 $\pm$ 2.5 & 89.0 $\pm$ 3.4 & 63.0 $\pm$ 3.8   \\
% & MedRax            & 50.0 $\pm$ 0.0 & 50.0 $\pm$ 0.0 & 1.0 $\pm$ 0.0   \\
\midrule
\multirow{4}{*}{Supervised Classifiers}
& RexNet-150        & 86.2 $\pm$ 3.9 & 87.8 $\pm$ 1.8 & 84.2 $\pm$ 5.5  \\
& RegNetY-1.6       & 86.4 $\pm$ 3.4 & 88.2 $\pm$ 4.4 & 84.4 $\pm$ 7.0   \\
& ResNet-50         & 86.6 $\pm$ 3.0 & 86.5 $\pm$ 4.3 & 86.8 $\pm$ 2.7   \\
& EfficientNet-v2   & 87.8 $\pm$ 3.0 & 88.8 $\pm$ 3.8 & 87.1 $\pm$ 8.4   \\
\midrule
\multirow{2}{*}{Ours}
& \textbf{AT-CXR (Rule-Based Router)}   & 93.8 $\pm$ 3.9 & 96.0 $\pm$ 3.6 & 92.9 $\pm$ 3.6   \\
& \textbf{AT-CXR (LLM Router)} &  \textbf{95.3 $\pm$ 3.3}  &  \textbf{96.7 $\pm$ 4.1}  &  \textbf{93.8 $\pm$ 2.8}    \\
\bottomrule
\end{tabular}%
}
\end{table*}

\begin{table*}[t]
\centering
\caption{Selective–prediction results on pulmonary–edema triage. We report AURC (area under the risk–coverage curve), the error rate among the most-confident 80\% of cases (risk@80\% coverage), and the maximal auto-acceptance rate under a 5\% error budget (coverage@5\% risk). Values are mean~\(\pm\)~standard deviation over five cross-validation folds using identical splits.}
\label{tab:selective}
\setlength{\tabcolsep}{6pt}
\footnotesize
\resizebox{\textwidth}{!}{%
\begin{tabular}{lccc}
\toprule
\textbf{Model} & \textbf{AURC (\%)(}\(\downarrow\)\textbf{)} & \textbf{risk@80\% cov. (\%)(}\(\downarrow\)\textbf{)} & \textbf{coverage@5\% risk (\%)(}\(\uparrow\)\textbf{)} \\
% \textbf{Model} & \textbf{AURC (\%)} & \textbf{risk@80\% coverage (\%)} & \textbf{coverage@5\% risk (\%)} \\
\midrule
RexNet-150                & 4.7 $\pm$ 1.8              & 9.1 $\pm$ 1.7               & 60.3 $\pm$ 15.5 \\
RegNetY-1.6               & 5.1 $\pm$ 0.8              & 8.1 $\pm$ 1.3               & 60.6 $\pm$ 10.0 \\
ResNet-50                 & 4.6 $\pm$ 1.3               & 7.5 $\pm$ 2.3               & 59.2 $\pm$ 21.5 \\
EfficientNet-v2           & 4.0 $\pm$ 0.6              & 6.5 $\pm$ 1.6               & 68.6 $\pm$ 13.8 \\
\textbf{AT-CXR (Rule-Based Router)}& 1.0$\pm$ 1.1               & 1.5 $\pm$ 1.7               &98.5 $\pm$ 2.2 \\
\textbf{AT-CXR (LLM Router)} & \textbf{0.9 $\pm$ 1.1}    & \textbf{1.4 $\pm$ 1.7}      & \textbf{98.9 $\pm$ 2.5} \\
\bottomrule
\end{tabular}
}
\end{table*}

\begin{table*}[t]
\centering
\caption{Per-image inference latency (mean, in seconds) for vision–language models (VLMs) and AT-CXR.}
\label{tab:latency-mean}
\setlength{\tabcolsep}{6pt}
\footnotesize
\resizebox{0.6\textwidth}{!}{%
\begin{tabular}{lc}
\toprule
\textbf{Model} & \textbf{Latency (s) (}\(\downarrow\)\textbf{)} \\
\midrule
GPT-4o               & 6.6 \\
GPT-4.1              & 5.7 \\
GPT-4.1-mini         & 4.6 \\
GPT-5                & 20.5 \\
GPT-5-mini           & 15.2 \\
MedGemma-4b          & 7.9 \\
MedGemma-27b         & 885.1 \\
CheXagent          & 3.8 \\
\midrule
\textbf{AT-CXR (Rule-Based Router)} & \textbf{1.5} \\
\textbf{AT-CXR (LLM Router)} & 3.3 \\
\bottomrule
\end{tabular}
}
\end{table*}

% \subsection{Implementation Details}
\section{Results and Discussion}
\noindent\textbf{Overall performance.}
Table~\ref{table:result} summarizes full-coverage (\(100\%\)) triage results. 
AT-CXR with LLM Router attains the highest Accuracy, Precision, and Recall, indicating superior end-to-end performance for pulmonary–edema triage. 
AT-CXR Rule-Based Router variant also performs strongly, ranking second across all methods in each metric. 
Among zero-shot VLMs, MedGemma-27b achieves the best accuracy, while among supervised classifiers, EfficientNet-V2 has the best performance. 
As expected, supervised classifiers outperform zero-shot VLMs on this task. 
We attempted to include LLaVA-Med and MedRAX; however, MedRAX predicted all cases as positive with 50\% accuracy, and LLaVA-Med failed to produce valid outputs reliably (empty JSON or prompt repetition). 

\medskip\noindent\textbf{Selective prediction.}
Table~\ref{tab:selective} reports selective–prediction metrics. 
AT-CXR with LLM Router obtains the lowest AURC (0.9\(\pm\)1.1\% vs.\ 4.0–5.1\% for supervised baselines), reflecting substantially better risk–coverage trade-offs. 
Moreover, it also attains the lowest error among the most-confident 80\% of cases (risk@80\% coverage = 1.4\(\pm\)1.7\%), representing a \(4{\times}\) – \(6{\times}\) error reduction compared with supervised classifiers (6.5–9.1\%). 
At a 5\% error budget, AT-CXR with LLM Router attains coverage@5\% risk of \(98.9\pm2.5\%\), meaning it could automatically resolve up to \(\sim99\%\) of cases while keeping the error within the auto-resolved subset \(\le 5\%\). This markedly exceeds the supervised baselines (59--69\%), with a gain of roughly 30 percentage points. Taken together, these results indicate that AT-CXR can confidently auto-resolve the vast majority of studies while maintaining a low error rate on the accepted subset.

\medskip\noindent\textbf{Latency.}
Table~\ref{tab:latency-mean} compares mean per-image latency. 
AT-CXR with Rule-Based Router has the lowest latency (1.5\,s), followed by AT-CXR with LLM Router (3.3\,s); both are faster than all evaluated VLMs, with some local VLMs being orders of magnitude slower. 
This efficiency arises because AT-CXR escalates only those cases that fail earlier checks (initial assessment, TTA stability, or MoE agreement), thereby avoiding VLM calls on easy cases. 
These latency advantages support the feasibility of the framework in settings where timely triage is critical.

\medskip\noindent\textbf{Design trade-offs.}
Contrasting the two router designs highlights a practical trade-off: the rule-based router offers the lowest latency (1.5\,s) with slightly lower accuracy (93.8\%), whereas the LLM router achieves the best accuracy (95.3\%) at modest additional latency (3.3\,s). 
Depending on clinical priorities (throughput versus maximal accuracy), a deployment can select the router variant that best aligns with operational constraints.

% needs \usepackage{booktabs}

\section{Future Work}
We will evaluate our framework for multi-label CXR triage. We will also rigorously assess the effect of medical-device suppression on routing decisions by providing device-suppressed images to the router rather than limiting suppression to a post-acceptance step, and we will quantify its impact on accuracy, calibration, and coverage–risk trade-offs.

% \section{Conclusion}
% In this paper, we proposed AT-CXR, an end-to-end uncertainty-aware agentic triage framework for Chest X-ray. In our proposed framework, we use out of domain (OOD) score and initial assessment confidence as a metrics to evaluate the uncertainty. Uncertainty guardril will automatically request router to double check these uncertain cases.
\section{Conclusion}
In this article, we proposed \textbf{AT-CXR}, an end-to-end, uncertainty-aware agentic framework for chest X-ray triage. The system fuses an out-of-distribution (OOD) score with initial-assessment confidence to form a unified uncertainty signal that governs actions and routing. A guardrailed policy with verification tools and optional escalation to a vision-language model yields automated outputs with auditable artifacts. We also explored two router designs, an LLM-based router and a rule-based router, which offer complementary operating points: the LLM-based version prioritizes accuracy at modest additional latency, whereas the rule-based version minimizes latency with a small loss in accuracy. Depending on clinical priorities, deployments can select the router variant that best aligns with throughput and accuracy requirements. Both variants outperform strong zero-shot VLMs and state-of-the-art supervised classifiers in full-coverage accuracy and in selective-prediction metrics, achieving lower AURC, lower error at 80\% coverage, and higher coverage at 5\% risk, with the lowest latency.

\begin{appendices}

\section{LLM Router Prompt}\label{AppA}
% \begin{tcolorbox}[breakable, colback=white, colframe=black!40,
                  % boxrule=0.4pt, sharp corners]
\begin{lstlisting}[style=router]
You are a safety-first controller for pulmonary-edema triage.
Goal: reach a final label (PE_yes / PE_no) with minimal risk, cost, and latency.
Available tools (router actions): accept, tta, moe, vlm, POST_ACCEPT.

HARD RULES (never violate; local guardrails may further restrict actions):
- Never auto-accept cases that are out-of-distribution (OOD) by the Mahalanobis threshold or with low confidence.
- Only ACCEPT when confidence is high and either the case is in-domain or TTA/MoE agrees.
- POST_ACCEPT is valid only if the case has already been ACCEPTed and final_label == "PE_yes".
  If chosen, the system will run post-accept utilities but NOT change the clinical label.

POST-ACCEPT UTILITIES (what POST_ACCEPT triggers; these are not separate actions):
- lung_seg, rib_seg_supp, device_seg_supp, LWI_calculate  (to compute a lung-whiteness index).
- CAM generation (class-activation map) for interpretability.

DECISION RUBRIC:
Pick exactly ONE next_tool from {accept, tta, moe, vlm, POST_ACCEPT}.
- TTA has the lowest cost and latency. 
- MoE has slightly higher cost and latency.
- VLM has the highest cost/latency and is treated as "abstain with suggested label".
- POST_ACCEPT may be used after a PE_yes acceptance to request the utilities above.
Return ONLY a JSON object with keys: next_tool, reason, stop, final_label, decided_by.

STATE:
{
  "p": <float>,                     // current PE probability
  "frd_maha": <float>,              // Mahalanobis distance
  "thr_maha": <float>,              // OOD threshold
  "tta_std": <float|null>,          // TTA posterior std if TTA ran, else null
  "moe_a": <float|null>,          // moe agreement rate if moe ran, else null
  "in_domain": <true|false>,
  "policy": {
    "mode": "<default|conservative|sensitive>",
    "p_accept": <float>,            // base accept threshold
    "std_ok": <float>,              // TTA agreement threshold
    "moe_thr": <float>,              // moe agreement threshold
    "max_auto_accept_frd_multiple": <float>
  },
  "already_ran": { "tta": <bool>, "moe": <bool>, "vlm": <bool> },
  "available_tools": ["accept"|"tta"|"moe"|"vlm"|"POST_ACCEPT", ...],
  "accepted": <true|false>,         // true only after an ACCEPT action has been taken
  "final_label": <"PE_yes"|"PE_no"|null> // label if already accepted, else null
}

Decide next_tool.
\end{lstlisting}
% \end{tcolorbox}

\section{VLM Prompt}\label{AppB}
We use the same instruction string for the VLM tool in our system and for all VLM baselines.

\begin{lstlisting}[basicstyle=\ttfamily\small,breaklines=true]
You are an image analysis agent for chest X-rays (CXR).
Return exactly ONE line, fields separated by '|', wrapped by markers:
===LINE===
<LABEL>|<CONF>|<EXPLANATION>
===END===
Definitions:
- CONF is a directional confidence in [0.0, 1.0]: 0.0 means definitely NO pulmonary edema; 1.0 means definitely YES pulmonary edema.
- LABEL must equal 1 if CONF >= 0.5, otherwise 0.
- EXPLANATION is a short phrase WITHOUT the '|' character.
Examples:
===LINE===
1|0.87|bilateral perihilar haze and interstitial markings
===END===
===LINE===
0|0.12|clear lungs without interstitial congestion
===END===
\end{lstlisting}

\end{appendices}

%%===========================================================================================%%
%% If you are submitting to one of the Nature Portfolio journals, using the eJP submission   %%
%% system, please include the references within the manuscript file itself. You may do this  %%
%% by copying the reference list from your .bbl file, paste it into the main manuscript .tex %%
%% file, and delete the associated \verb+\bibliography+ commands.                            %%
%%===========================================================================================%%

\bibliography{sn-bibliography}% common bib file
%% if required, the content of .bbl file can be included here once bbl is generated
%%\input sn-article.bbl

\end{document}